\documentclass[]{raa}            
\usepackage{graphicx,times}
\usepackage{natbib}

\providecommand{\bm}[1]{\mbox{\boldmath$#1$\unboldmath}}

\begin{document}

   \title{Neural Network Prediction of solar cycle 24
}

 \volnopage{ {\bf 2010} Vol.\ {\bf 9} No. {\bf XX}, 000--000}
   \setcounter{page}{1}

   \author{A. Ajabshirizadeh
      \inst{1}
   \and N. Masoumzadeh Jouzdani
      \inst{1,3}
   \and S. Abbassi \mailto{}
      \inst{2,3}
   }


   \institute{Department of Physics, University of Tabriz, Tabriz, Iran\\
    \email{abbassi@mail.ipm.ir}
        \and
             School of Physics, Damghan University, P. O. Box 36175-364 Damghan, Iran\\
        \and
             School of Astronomy, Institute for Research in Fundamental Sciences (IPM), Tehran, Iran\\
\vs \no
   {\small Received [year] [month] [day]; accepted [year] [month] [day] }
}

\abstract{ The ability to predict the future behavior of solar activity has become of extreme
importance due to its effect on the near Earth environment. Predictions of both
the amplitude and timing of the next solar cycle will assist in estimating the various
consequences of Space Weather. The level of solar activity is usually expressed by international sunspot number ($R_z$). Several prediction techniques have been applied and have achieved varying degrees of success in the domain of solar activity prediction. In this paper, we predict a solar index ($R_z$) in solar cycle 24 by using the neural network method. The neural network technique is used to analyze the time series of solar activity. According to our predictions of yearly sunspot number, the maximum of cycle 24 will occur in the year 2013 and will have an annual mean sunspot number of 65. Finally, we discuss our results in order to compare it with other suggested predictions.
}

   \authorrunning{ A. Ajabshirizadeh, N.Masoumzadeh\& S. Abbassi }            
   \titlerunning{Neural Network Prediction of solar cycle 24 }  
   \maketitle


%
%
\section{Introduction}           
\label{sect:intro}

The successful prediction of a future event is arguably the
most powerful way of confirming a scientific theory. Commonly in
physics, a theory that is describing a system in a natural world is
regarded as correct and therefore useful if it can use the state of
the system at one time to reconstruct the state of the system at
some other time, in past or future.

The prediction of solar activity for a few years is the
oldest problem in solar physics, arising as soon as solar cycle
itself was discovered. Unfortunately, this problem has not been
solved, probably because the series of observational data available
are not long enough for purely statistical analysis, and because we
do not quite understand the physical nature of this phenomena.

Most of the space weather phenomena are influenced by variations in
solar activity. During the years of solar maximum there are more
solar flares causing significant increase in solar cosmic ray
intensity. The high-energy particles disturb communication systems
and affect the lifetime of satellites. Coronal mass ejections and
solar flares are the origin of shocks in solar wind and cause
geomagnetic disturbances in the earth's magnetosphere. The high rate
of geomagnetic storms and sub-storms results in atmosphere heating
and drag of Low Earth Orbit (LEO) satellites. Solar activity
forecasting is especially useful to space mission centers as in the
orbital trajectory parameters of satellites are greatly affected by
variations of  solar activity. A dramatic effect, not only on the
Earth's upper atmosphere, disturbing the orbits of satellites, but
also on power grids on the ground, e.g. the power cuts in Quebec,
Canada in 1989. The level of solar activity is usually expressed by the
Zurich or International sunspot number.

Although the solar activity presents some clear periodicities, its
prediction is quiet difficult but not impossible, as a large range
of forecasting methods using predict the occurrence and amplitude of
solar cycle is categorized to two models; statistical models and
physical models. In statistical models, it is usual to represent the
evolution of a physical system by using a time series. In Contrast
with a physical model, the statistical model only attempts to
explain the system, and in particular a time  series associated with
it, in terms of itself, and perhaps in terms of correlation with
other time series associated with the system. At this point, it is
appropriate to address a common concern, which for obvious reasons
is most usually expressed by physicists: what reasons are there for
constructing a model that contains no physical understanding? Here
are three reasons. Firstly, simply writing down the data as a time
series, together with organizing and examining it, is the first step
in the scientific method: analyzing the sequence as a time series
governed by a statistical model is a natural first step, until such
time as a physical theory can be formulated. The second reason is
that predictions from a statistical model might simply be useful in
their own right. For example, in day to day life it makes no
difference to most people whether the weather forecast was made from
a statistical model or from a physical one. The final and most
important reason is that it might be impossible for the physical
system to be predicted from the basic physical principles governing
it. This can be because the system is simply too complicated, which,
for example, is the case for a plasma (Conway~\cite{con98}).

One of the statistical models using for predicting the data is
artificial neural networks method. The use of artificial neural networks
has been recognized recently as a promising way of making predictions on temporal series with chaotic
or irregular behavior (Weigend~\cite{wei90}). This technique has already been applied in the framework
of solar-terrestrial physics for prediction of geomagnetic induced current and storms (Lundstedt~\cite{lun92}) and as a way of recognizing a pattern in the onset of a new sunspot cycle (Koons~\cite{koo90}).

The aim of this paper is to predict the solar cycle. The structure of the paper is as follow. In section
2 we provide a brief summary of the neural network methodology
employed. In section 3 we introduce the results of our network
architecture to generate our best estimate of the behavior of cycle
24, and In section 4, the conclusions and their comparison are presented.


\section{Artificial Neural Network}
\label{sect:ANN}

An artificial neural network (ANN) is an information-processing system consisting of a large number of simple processing elements called neurons or units.
The Neural Network (NN) system is characterized by (i) its pattern of connection between the neurons, (ii) its method of determining the weights on the connections (training or learning algorithm) and (iii) its activation function.
In other words ANNs  are parallel computing systems that are widely used in prediction, pattern recognition and classification. Neural networks with sufficient number of hidden units
can approximate any nonlinear function to any degree of accuracy (Hornik et al.~\cite{hor89}).

There exits various types of NN; however, for our predictions, we have used the most popular and simple NN,
which is the Feed Forward Neural Network (FFNN) employing the Levenberg-Marquardt of errors
learning algorithm ( Levenberg~\cite{lev44}, Marquardt~\cite{mar63} ). Although back-propagation of
errors learning algorithm ( Rumelhart~\cite{rum86}) is more famous and usually use in FFNN, it is also known as an algorithm with a very poor convergence rate. More significant improvement was possible by using various second order approaches
such as Newton, conjugate gradient, or the Levenberg-Marquardt (LM) method. The LM algorithm is now considered as the most efficient. It combines the speed of Newton algorithm
with the stability of the steepest decent method (Hagen et al. ~\cite{ha94}).

In a FFNN arrangement neurons (units) between layers are connected in a forward direction. Neurons in a given layer do not connect to each other and do not take inputs from subsequent layers. The input units send the signals to the hidden units, which then process the received information and pass the results to output units. The output units produce the final response to the inputs signals.
A database of historical data describing the relation ship between a set of inputs and known outputs is used to define the inputs and output units.
Feed Forward networks often have one or more hidden layers of sigmoid neurons followed by an output layer of linear neurons.
Multiple layers of neurons with nonlinear activation functions allow the network to learn nonlinear and linear relationships between input and output vectors.
The linear output layer lets the network produce values outside the range -1 to +1.
\begin{figure}    
   \centerline{\includegraphics [width=0.5\textwidth,clip=]{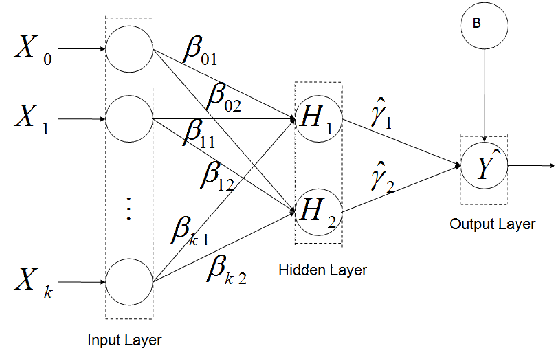}
              }
              \caption{A  FFNN with one hidden layer and  one output.}
   \label{F-simple}
\end{figure}

A typical FFNN is depicted as follows:
Algebraic form of neural network can be written
\begin{equation}
Y=y_{0}+\sum_{j=1}^{h}y_{j}f_{j}(X,w_{j})
\end{equation}

Where $w_{j}$ is the vector of weights for $jth$ neuron , $X=(X_{1},X_{2},..., X_{k})$ is vector of explanatory variables, and $f_{j}(X,w_{j})=G(w_{0j}+X'w_{j}), j=1,...,h$ shows output of the hidden unites. The function $G$ is any
activation function such as hyperbolic tangent function $G(n)=(e^{n}-e^{-n})(e^{n}+e^{-n})^{-1}$ or logistic function $G(n)=(1+e^{-n})^{-1}$. An FFNN can compose from more than one hidden layer as well it can has multi output which is similar to system of nonlinear regression equations.

The FFNN is organized here with three layers; input, hidden, and output layers.
The activation function in the first layer is log-sigmoid, and the output layer activation function is linear.Units between layers are connected by weights that are optimized for a minimum of the root mean square error(RMSE)between a known output and the predicted output.
Training is the process by which the weights are adjusted according to  Levenberg-Marquardt algorithm. A simplified definition of a NN is computer program that has been trained to learn the relationship between a given set of inputs and a known outputs.
In general, training a NN requires an optimum network architecture and sufficient historical information about the time series.
The architecture of a feed-forward network is specified by the number of neurons used in the input, hidden and output layers of the network. The input layer needs sufficient number of neurons so that the network has access to enough of the recent history of the time series.
The hidden layer of the network is responsible for the nonlinear processing capability of the network and as such needs to have sufficient neurons to represent the underlying complexity of the time series. We only consider networks with one output, which is required to produce a prediction a number of years ahead of the must recent input.
NNs are trained until the RMSE between the output values predicted by the network and the target output values has reached a minimum. At this point we say that the optimum result has been reached for the given situation. As applied to Sunspot Number (SSN) prediction, the RMSE was defined as

\begin{equation}
RMSE=\sqrt{\frac{1}{N}\sum_{i=1}^N(SSN_{obs}-SSN_{pred})^2}
\end{equation}

where N is the number of training patterns, $SSN_{obs}$ and $SSN_{pred}$ are the observed and predicted (SSN) values. Generally, the time series is split into two data sets: a training set and a testing set.
The training set is used to adjust the weights during training, while the testing set is used to verify the prediction performances of the network.
Neural networks with large numbers of parameters are more in risk of overfitting. Overfitting is the problem of very bad predictions for the out of sample data in spite of  very good results for in sample data.
Here, for overcoming this problem we used early stopping method which stops training when validation set fails to reduce validation sample RMSE (Baum~\cite{bau89}).

\begin{table}
\caption{ Predicted values of sunspot number from 2008 to 2018
}
\label{T-simple}
\begin{tabular}{lcclcclcc}     
\hline                   
      &          &           &     &          &         &      &          &     \\
 Year & Predicted & min &Year & Predicted  & min & Year & Predicted  & min  \\
      & values   & RMSE      &     &values &RMSE &      & values     & RMSE    \\

  \hline

1986& 12.68~ & 0.060  & 1997  & 12.40~ & 0.055 & 2008 & 14.46~  & 0.039 \\
1987& 36.99~ & 0.064  & 1998  & 66.74~& 0.053  & 2009 & 16.23~  & 0.037 \\
1988& 86.17~ & 0.057  &1999   &114.68 & 0.051  & 2010 & 17.91~  & 0.049 \\
1989&144.80  & 0.060  &2000   &132.90 & 0.056  & 2011 & 43.50   & 0.045 \\
1990&135.97  & 0.055  &2001   & 115.51 & 0.060 & 2012 & 57.64   & 0.047 \\
1991&124.08  & 0.058  &2002   & 104.19 & 0.055 & 2013 & 65.43   & 0.045 \\
1992& 92.14~ & 0.053  &2003   & 64.75~&  0.051 & 2014 & 56.74~  & 0.042 \\
1993& 57.79~ & 0.053  & 2004  & 42.20~&  0.051 & 2015 & 48.37~  & 0.042 \\
1994& 38.55~ & 0.051  & 2005  & 27.37~&  0.043 & 2016 & 18.58~  & 0.041 \\
1995& 20.63~ & 0.046  & 2006  & 19.94~&  0.039 & 2017 & 10.82~  & 0.040 \\
1996& 11.22~ & 0.052  & 2007  &  15.40~& 0.042 & 2018 & 14.17~  & 0.041\\

  \hline
\end{tabular}
\end{table}

Finally, before going on to present our results, we mention two different ways in which neural networks can be used to produce predictions. Firstly, in what we turn "direct prediction",
the network only relies on actual known data to generate any predictions. Consequently the furthest ahead prediction obtainable is limited by the last known data point in the time series plus the predict ahead time of the individual network.
Alternatively, networks can be used to predict iteratively, sometimes called multistep prediction, in which the networks' predictions are subsequently fed back into the input layer as new data points.
Potentially this allows networks to predict arbitrarily far ahead; in practice, as predicted values make up more and more of the supposedly known input data, errors can be compounded recursively until no estimate of accuracy of the results can be calculated (Conway~\cite{con98}).

\section{Solar indices forecasting}
\label{sect:data}

Here we wish to predict up to 10 years ahead, and consequently, we use yearly sunspot number since the use of monthly data would require potentially large network. Furthermore, Hoyt et al.~\cite{hoy94} have shown that some of Wolf's reconstructed values were wrong, particularly for the early cycles 1-7. Thus only the post-1850 data can be considered wholly reliable. It should be considered that cycle 23 began in 1996 May and reached its maximum in 2000 April, and now it is inferred to end in 2008 December (or probably later); therefore, its length should be 12.6 yr (or longer)(Li~\cite{li09}).The sunspot number yearly mean value were obtained electronically from the website:ftp://ftp.ngdc.noaa.gov/STP/SOLAR\_DATA/.

Regarding time difference between data accessing, we choose various network architecture for our time series. After a massive work of trial and error, We process sunspot number time series with a neural network of 128-42-1 structure which means we used the sunspot values for the years 1882-2009 as the training set.

For the sunspot number $ R_z$, we obtained 2013 as the year of next maximum with a value of around 65. Regarding the accuracy of the year of maximum prediction, for the two cycles predicted with this network, in two cases the date of maximum was predicted correctly. With comparison of predicted value and observed value of Solar Cycles (SCs) 22 and 23, Uncertainty about the value of the sunspot maximum have been obtained $\pm13$.
All of the predicted values for sunspot number have been added to Table 1. Also, comparison between the 1986-2009 observed sunspot number and the predictions of neural networks are shown in figure 2 as well as the predicted shape and amplitude of SC 24 in terms of yearly sunspot number.

\begin{figure}    
   \centerline{\includegraphics [width=0.5\textwidth,clip=]{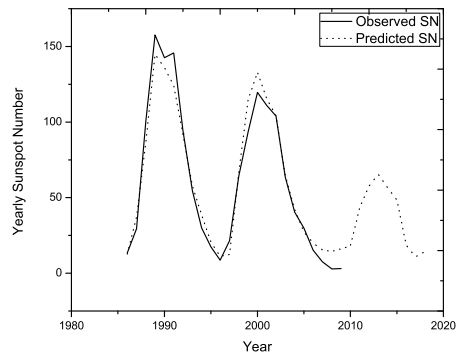}
              }
              \caption{Observed SCs 22 and 23 (solid line) and the predicted SCs 22,23 and 24 (dashed line)
              in terms of yearly mean sunspot numbers.   }
   \label{F-simple}
   \end{figure}

\section{Discussion and conclusions}
\label{sect:discussion}

Our neural network method is based on one hidden layer. For having reliable result, we use multi-step prediction to have only one reasonable output. In terms of processing data, by changing back-propagation algorithm to Levenberg-Marquardt algorithm our feed-forward neural network model becomes faster since LM algorithm speeds up convergence while limiting memory requirements (Battiti~\cite{bat92}).
We saw almost a similarity between predicted Solar cycle 24 and Solar cycle 20 . We predict a SC 24 with a maximum of  $65\pm13$ occurring in 2013. In general, our result is close to other prediction made for solar cycle 24. For example, Li et al.~\cite{li05} obtained 2013 a maximum of cycle 24 with statistical method. Also a recent article by Wang et al.~\cite{wang09}, using similar descending phases and a cycle grouping, predicted that peak amplitude for that monthly smoothed sunspot number in the solar cycle 24 is near $100.2\pm7.5$, occurring in 2012. Furthermore, Chumak et al.~\cite{chumak10} predicted that the maximum amplitude of cycle 24 is $90\pm20$ which is in agreement with our results.
Finally, Our prediction fits well within the limits of the others as indicated in Pesnell~\cite{pes08} where an average cycle was predicted using other methods such as statistical and precursor methods.

\label{lastpage}

\end{document}